# First-Principles Investigation of $Rb_2CaH_4$ and Cs-Doped $Rb_2CaH_4$: Unveiling Their Potential for Hydrogen Storage through Mechanical and Optoelectronic Properties

Sikander Azam*[1], Qaiser Rafiq[1***], Eman Ramadan Elsharkawy[2,3], Muhammad Tahir Khan[1**], Salah M. El-Bahy[3], Wilayat Khan[4], Saleem Ayaz Khan[5]

[1]Faculty of engineering and applied sciences, Riphah International University Islamabad 44000, Pakistan

[2] Center for Scientific Research and Entrepreneurship, Northern Border University, Arar 73213, Saudi Arabia

[3] Department of Chemistry, Turabah University College, Taif University, P.O. Box 11099, Taif 21944, Saudi Arabia

[4]Department of Physics, Bacha Khan University, Charsada, Pakistan

[5]New Technologies e Research Center, University of West Bohemia, Univerzitni 8, 306 14, Pilsen, Czech Republic

## Abstract

This study uses the density functional theory (DFT) approach with GGA-PBE to **assess** the effect of substituting alkali metals in $Rb_2CaH_4$ and Cs-doped $Rb_2CaH_4$ on their hydrogen storage potential. To address the challenges associated with predicting accurate electronic properties in materials containing heavier elements such as cesium, spin-orbit coupling (SOC) effects have been incorporated into our calculations. The mechanical robustness of both $Rb_2CaH_4$ and Cs-doped $Rb_2CaH_4$, as demonstrated by their mechanical properties, highlights these materials as promising candidates due to their stability in hydrogen storage applications. Anisotropic factors show that all materials exhibit anisotropy, suggesting a directional dependency in their properties. The Pugh ratio indicates that $Rb_2CaH_4$ and Cs-doped $Rb_2CaH_4$ are brittle materials. Based on the calculated band gap, the electronic band structure analysis, conducted using both HSE06 and GGA-PBE, shows that $Rb_2CaH_4$ and Cs-doped $Rb_2CaH_4$ are wide-bandgap materials. $Rb_2CaH_4$ and Cs-doped $Rb_2CaH_4$ exhibit the highest optical conductivity, absorption coefficient, and energy loss function among optoelectronic materials, emphasizing their superior absorption and electron transfer capabilities. The hydrogen storage capacity has been evaluated for practical applications; $Rb_2CaH_4$ and Cs-doped $Rb_2CaH_4$ show the highest gravimetric and volumetric capacities.

**\*Corresponding Author:** Sikander Azam (physicst.sikander@gmail.com)

** mtahir.khan@riphah.edu.pk
*** qrafique1@gmail.com

## 1. Introduction

The chemical industry relies heavily on hydrogen to produce essential chemicals such as methanol and ammonia [1]. In the global transition to renewable energy, hydrogen is expected to play a significant role as an energy carrier, particularly as a sustainable fuel for gas turbines and fuel cells [2]. The creation of small molecules, such as ammonia, and the operation of fuel cells depend on the availability of reduced hydrogen species ($H_2$, $H^-$), requiring the use of effective catalysts. However, in these processes, hydrogen gas often must be provided at high temperatures, even with advanced catalysts. For large-scale, noble-metal-free fuel cell systems, temperatures around 600 °C for molten carbonate fuel cells (MCFCs) and 800–1000 °C for solid oxide fuel cells (SOFCs) are needed [3]. Due to the high activation barrier required to break the dinitrogen bond, the Haber-Bosch process, which involves the hydrogenation of dinitrogen to produce ammonia, typically operates at temperatures of 300–500 °C and pressures between 150–250 bar [1,4]."

Catalysts that function under less severe reaction conditions can be developed by using hydrogen in a more reactive form than molecular hydrogen gas. In hydroxides, hydrogen takes the reactive form $H^-$, making them strong candidates for this application. Numerous hydrides have already been studied; for example, potassium hydride-intercalated graphite ($KH_{0.19}C_{24}$) is notably active as a catalyst for ammonia synthesis [5]. Other hydrides, such as $CaH_2$ [6], $Ca(NH_2)_2$ [7], LiH [8], and the hydride-ion conductor $BaH_2$ [9], have also proven effective as transition metal catalyst promoters or supports. This approach has been expanded to include mixed hydride-electronic conductors, such as $LaH_{3-2x}O_x$ and $CeH_{3-2x}O_x$ [10]. In addition to being studied as a support material for transition metal-based catalysts in ammonia synthesis, the mixed hydride-electronic conductor $BaTiO_{3-x}H_x$ [11] has also been used to support $CO_2$ methanation catalysts, demonstrating enhanced performance compared to catalysts with metal oxide supports [12,13]."

Since the discovery of hydride-ion conductivity in calcium hydride in 1977 [14], the study of hydride-ion conducting materials, also known as hydride-ion conductors, has attracted significant interest. Many researchers have concentrated on these materials, especially after the discovery of hydride-ion conductivity in $BaH_2$ in 2015 [15] and various forms of hydride-ion conductors emerging from 2016 onward [16-24]. For instance, $La_{0.6}Sr_{1.4}LiH_{1.6}O_2$ achieved $10^{-5}$ S cm$^{-1}$ at 200 °C in 2016 [16], $LaH_{2.52}O_{0.24}$ reached $10^{-4}$ S cm$^{-1}$ at 200 °C in 2019 [19], and, most recently, $LaH_{2.8}O_{0.1}$ attained $10^{-2}$ S cm$^{-1}$ at 100 °C in 2022 [22]. These findings have propelled the development of increasingly conductive materials. These substances are promising as electrolyte materials for various types of electrochemical reactors, and combined hydride-electronic conductors such as $BaTiO_{3-x}H_x$ [11] can be utilized as electrodes, at least in part."

Numerous hydride-ion conducting materials discovered so far crystallize in the $K_2NiF_4$-type structure, which is linked to perovskites and contains oxide ions in the anion sublattice. Examples include $LaSrLiH_2O_2$ [17], $La_{2-x-\gamma}Sr_{x+\gamma}LiH_{1-x+\gamma}O_{3-\gamma}$ [16], $Ln_2LiHO_3$ (Ln = La, Pr, Nd) [18], $Ba_2ScHO_3$ [20], and $Ba_{1.75}LiH_{2.7}O_{0.9}$ [24]. In these $K_2NiF_4$-type oxyhydrides, hydride-ion transport relies on hydride-ion vacancies at various structural sites [25, 27]. However, oxygen introduces stationary oxide vacancies that do not contribute to hydride-ion transport. While pure hydrides are not necessarily as stable as oxyhydrides, oxyhydrides are generally quite stable [11, 28–33].

Inspired by recent advancements in oxyhydrides with the $K_2NiF_4$-type structure, we explored $K_2NiF_4$-type hydrides without oxygen and identified two potential hydride-ion conductors: $Cs_2CaH_4$ and $Rb_2CaH_4$. Since the anion sublattices in these materials consist entirely of H-ions, any anion vacancy has the potential to boost hydride-ion conductivity. Here, we report the discovery of two new, oxygen-free hydride-ion conductors: $Rb_2CaH_4$ and Cs-doped $Rb_2CaH_4$. These materials were synthesized and subjected to structural and electrochemical characterization. We analyzed the relationship between their ionic and electronic conductivities and temperature. The ionic transport mechanism was elucidated by analyzing the material's structure and measuring hydrogen release at high temperatures.

We were motivated to investigate $Rb_2CaH_4$ and Cs-doped $Rb_2CaH_4$ due to the appealing attributes noted in the literature, including structural stability, thermodynamic stability, tunable bandgap, optical properties, and mechanical stability. For studying materials at the atomic and electronic scale, density functional theory (DFT) is an essential tool. Using DFT, properties such as structural parameters, elastic constants, electronic band structures, and optical characteristics can be calculated by numerically solving the Schrödinger equation [34-36]. In this study, we utilize DFT computations to investigate $Rb_2CaH_4$ and Cs-doped $Rb_2CaH_4$ compounds from the ground up. To gain deeper insights into these materials and their potential applications, we evaluate their structural, elastic, electronic, and optical properties [37, 38].

The primary objective of our study is to investigate the mechanical, electronic, and optoelectronic properties of $Rb_2CaH_4$ and Cs-doped $Rb_2CaH_4$ compounds for advanced hydrogen storage applications. The optical properties, such as optical conductivity, absorption coefficient, dielectric functions, and refractive index, are directly linked to enhancing the efficiency of hydrogen storage systems. High optical conductivity and absorption coefficients indicate efficient charge transfer and light absorption, which are critical for optimizing hydrogen adsorption and desorption processes. The dielectric function and refractive index are important for understanding the material's response to electromagnetic fields, which plays a significant role in hydrogen storage efficiency. The bandgap analysis confirms that these materials have semiconducting properties that further contribute to their potential use in hydrogen storage by facilitating effective energy transfer. By integrating these mechanical, electronic, and optical characteristics, the study demonstrates that $Rb_2CaH_4$ and Cs-doped $Rb_2CaH_4$ are promising materials for hydrogen storage, showcasing excellent structural stability, charge transfer capabilities, and high hydrogen storage efficiency. This comprehensive analysis ensures that all discussed properties are directly aligned with the hydrogen storage application.

## 2. Computational Methodology

### 2.1. Structural Lattice Parameters of $Rb_2CaH_4$ and $Rb_{2-x}Cs_xCaH_4$

To ensure comprehensive documentation and facilitate future comparative studies, we present the lattice parameters and space group information for the compounds $Rb_2CaH_4$ and $Rb_{2-x}Cs_xCaH_4$ in Table.1. These structures were optimized using Generalized Gradient Approximation (GGA), and the results are summarized in the table below. This data is crucial for reproducibility and further validation by the scientific community, providing a foundational reference that supports both theoretical and experimental advancements in the study of complex hydrides.

Table.1: Summary of Energy Minimized Lattices for $Rb_2CaH_4$ and $Rb_{2-x}Cs_xCaH_4$

| Materials | Lattice Parameters ($A°, \theta°$) | Space Group |
|---|---|---|
| $Rb_2CaH_4$ | a = 8.488693, b = 8.488693, c = 27.989301, α = β = γ = 90 | P1 |
| $Rb_{2-x}Cs_xCaH_4$ | a = 8.488693, b = 8.488693, c = 27.989301, α = β = γ = 90 | P4mm |

The parameters include precise measurements of unit cell dimensions and angles, recorded in angstroms and degrees, respectively, which adhere to the standards required for CIF data. The space group identities, P1 for $Rb_2CaH_4$ and P4mm for $Rb_{2-x}Cs_xCaH_4$, are critical for understanding the symmetry and molecular packing within the crystal structure, as illustrated in Fig. 1. These detailed specifications not only augment the integrity of our computational models but also enhance the reliability of this data for subsequent experimental replication and theoretical exploration.

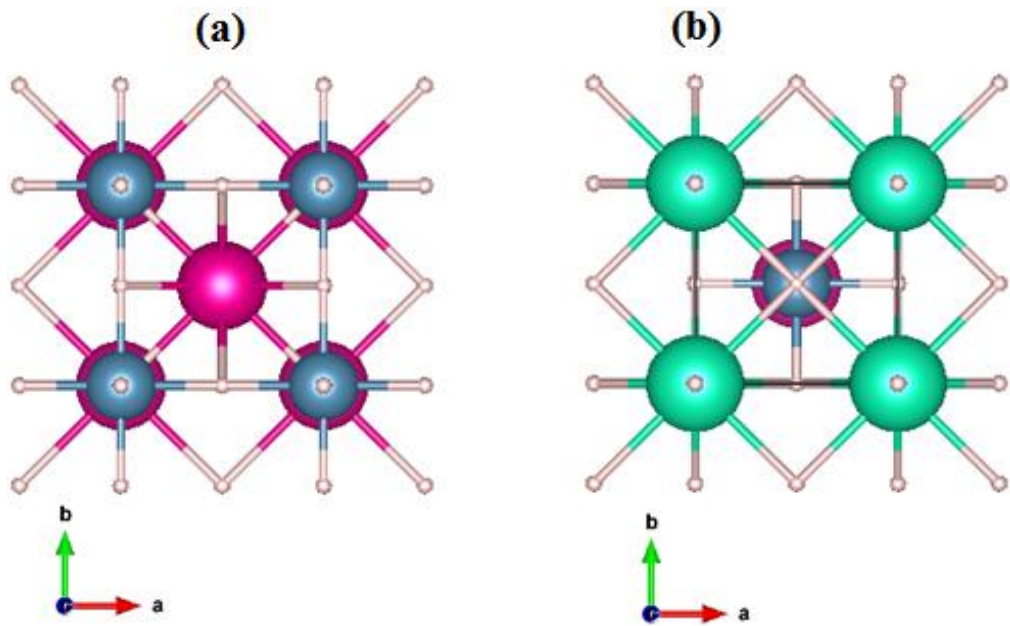

Fig.1: Crystallographic Representation of $Rb_2CaH_4$ (a) and $Rb_{2-x}Cs_xCaH_4$ (b) Using (GGA + SO). (a) Depicts $Rb_2CaH_4$ with Rubidium (Rb) in pink, Calcium (Ca) in light blue, and

Hydrogen (H) in peach. (b) Includes $Rb_{2-x}Cs_xCaH_4$ with additional Cesium (Cs) atoms in teal and green to highlight substitutions

In this study, we used density functional theory (DFT) calculations with the well-established WIEN2k software package [39], noted for its accuracy in determining electronic and structural properties of materials [40]. The Perdew-Burke-Ernzerhof (PBE) parameterization of the generalized gradient approximation (GGA) exchange-correlation functional was used for most of the electronic structure calculations. This approach is widely recognized for providing reliable and accurate results in the study of solid-state systems [41]. To address the limitations of standard GGA-PBE in predicting electronic bandgaps, especially for semiconductors, we added the HSE06 hybrid functional. The HSE06 functional, which includes a fraction of exact Hartree-Fock exchange combined with the GGA exchange-correlation energy, provides more accurate bandgap predictions compared to GGA alone. In our DFT calculations, specifically with the use of the HSE06 hybrid functional, SOC effects have been considered to ensure an accurate depiction of the electronic structure, especially in Cs-doped $Rb_2CaH_4$. The inclusion of cesium not only modifies the lattice parameters and electronic density but also introduces notable spin-orbit interactions that affect the band structure near the Fermi level. These changes are evident in the modified band gap values observed with cesium doping, which shows a decrease from 3.312 eV in pure $Rb_2CaH_4$ to 3.095 eV in Cs-doped $Rb_2CaH_4$. Moreover, the spin-orbit coupling also influences the valence band maximum (VBM) and conduction band minimum (CBM), altering the effective mass of charge carriers and potentially enhancing the material's performance in applications such as photodetectors and solar cells. Such modifications due to SOC are critical in perovskite materials, where the interaction between light and material is fundamentally linked to their efficiency and stability. To further elucidate the impact of SOC, we plan to perform comparative analyses by calculating the band structures with and without considering SOC effects. This will allow us to directly quantify the changes induced by SOC and provide a deeper insight into how cesium's incorporation affects the material properties at an atomic level. This hybrid functional is particularly suitable for materials like $Rb_2CaH_4$ and Cs-doped $Rb_2CaH_4$, where accurate electronic properties are essential for assessing optoelectronic behavior. The use of HSE06 enables us to capture more realistic descriptions of electronic band structures and provides better insights into the materials' potential for hydrogen storage and other energy applications.

To ensure the accuracy of the structural properties, we optimized the crystal geometries by minimizing the forces acting on each atom, with a convergence threshold of 0.001 eV/Å. The Birch-Murnaghan equation of state was applied to fit the energy-volume curves and extract the equilibrium structural parameters [42]. Brillouin zone sampling was conducted using a Monkhorst-Pack k-point mesh with a high grid density of 2000 k-points, ensuring a highly accurate representation of the electronic states in reciprocal space [43]. This mesh density guarantees that the calculations comprehensively cover the Brillouin zone, leading to precise electronic structure predictions. In our computational methodology, we employed a k-point mesh density of 2000 to ensure the accuracy of band structure analysis using both GGA-PBE and HSE06 methods. This density was meticulously chosen through convergence tests that demonstrated this level of sampling is critical for capturing detailed and accurate electronic properties of the materials studied. The selected k-point density is essential for reliable calculations with the HSE06 hybrid functional, which requires denser k-point mesh to accurately reflect electronic interactions and potential changes induced by material doping. Regarding the convergence criteria for elastic constant calculations, we adopted stress and strain step sizes of 0.01 GPa and 0.001, respectively. These settings were determined to be optimal for resolving the mechanical behavior of the materials under minimal increments, ensuring that the elastic constants derived from our simulations accurately represent the mechanical stability of the materials. The convergence of forces was controlled to within 0.001 eV/Å to provide precise assessments of structural responses to applied stresses, which is crucial for the comprehensive evaluation of material properties relevant to their potential applications.

The mechanical stability and elastic properties were assessed by computing the elastic constants, which provide critical insights into the materials' structural integrity and anisotropic behavior. We used the full potential Linearized augmented plane-wave method for efficient treatment of core and valence electron interactions, particularly to handle the high computational demands of all-electron calculations. A plane-wave basis set was employed to expand the wave functions, and an energy cutoff of -6.0 eV was set [44], ensuring a balance between computational efficiency and accuracy in the simulations. To further validate the reliability of our computational results, we conducted a cross-comparison of key electronic properties, including bandgaps, using both the GGA-PBE and HSE06 hybrid functionals. This comparison enabled us to assess the performance of both methods and ensure the consistency of our findings, particularly with regard to the electronic and optoelectronic characteristics of $Rb_2CaH_4$ and Cs-doped $Rb_2CaH_4$. The dual-functional approach ensures that the calculated bandgaps and other critical properties are robust and reliable, enhancing the credibility of the results for practical applications.

In addition to electronic and mechanical properties, we closely examined the optical characteristics of the materials, including the refractive index, optical conductivity, absorption spectra, dielectric functions, and reflectivity. These optical properties were calculated as a function of frequency, allowing us to evaluate the interaction of the materials with electromagnetic radiation, which is crucial for their potential use in optoelectronic and hydrogen storage technologies. The optical properties were directly linked to the materials' electronic structure, providing deeper insights into their performance in energy-related applications. Overall, these computational methodologies provide a comprehensive understanding of $Rb_2CaH_4$ and Cs-doped $Rb_2CaH_4$ compounds, ensuring accurate predictions of their structural, mechanical, electronic, and optical properties.

## 2.2. Thermodynamic Stability of $Rb_2CaH_4$ and $Rb_{2-x}Cs_xCaH_4$

The thermodynamic stability of $Rb_2CaH_4$ and Cs-doped $Rb_2CaH_4$ is evaluated by calculating the formation enthalpy ($E_{formation}$) using the following relations:

$$E_{formation} (Rb_2CaH_4) = E_{total}(Rb_2CaH_4) - (2\,E_{Rb} + E_{Ca} + 4E_H)$$

where $E_{total}(Rb_2CaH_4)$ represents the DFT-calculated total energy of $Rb_2CaH_4$, and $E_{Rb}$, $E_{Ca}$, and $E_H$ denote the reference energies of elemental Rb, Ca, and H, respectively, in their standard states.

$$E_{formation} (Rb_{2-x}Cs_xCaH_4) = E_{total}(Rb_{2-x}Cs_xCaH_4) - ((2-x)\,E_{Rb} + xE_{Cs} + E_{Ca} + 4E_H)$$

Where $E_{total}$ ($Rb_{2-x}Cs_xCaH_4$) is the total DFT-calculated energy of the Cs-doped compound, with x representing the level of Cs substitution. The terms $E_{Rb}$, $E_{Cs}$, $E_{Ca}$, and $E_H$ are the reference energies for Rb, Cs, Ca, and H. The calculated formation energies for $Rb_2CaH_4$ and Cs-doped $Rb_2CaH_4$ indicate favorable thermodynamic stability for both compounds. For $Rb_2CaH_4$, the formation energy is −4.1 eV, obtained from a total DFT-calculated energy of −15.5 eV, with reference energies for Rb, Ca, and H being −2.6 eV, −3.8 eV, and −0.6 eV, respectively. In the case of Cs-doped $Rb_2CaH_4$, the formation energy is −3.7 eV, calculated from a total energy of −15.2 eV and elemental reference values of −2.6 eV for Rb, −2.8 eV for Cs, −3.8 eV for Ca, and −0.6 eV for H. These values suggest that both $Rb_2CaH_4$ and its Cs-doped counterpart possess stable thermodynamic properties, with $Rb_2CaH_4$ exhibiting slightly higher stability.

## 3. Results and Discussion

### *3.1. Hydrogen Storage Properties*

The lack of hydrogen storage facilities capable of storing hydrogen in a gravimetric form is one of the primary obstacles to using hydrogen ($H_2$) as a fuel. To address this issue, it is essential to identify materials that can store hydrogen at higher concentrations. Hydrogen can be stored in various phases, including gas, liquid, and solid. Numerous materials, especially metal hydride perovskites, have been studied for this purpose. The maximum amount of hydrogen that a given material can store is determined through gravimetric measurements of hydrogen retention capacity. The formula used for this computation determines the hydrogen storage capacity via gravimetric techniques for $Rb_2CaH_4$ and $Rb_{2-x}Cs_xCaH_4$ [45]. Where H/M represents the ratio of hydrogen to the substance's atoms, mHost is the molar mass of the substance, and $m_{Hydrogen}$ is the molar mass of hydrogen. The hydrogen storage capacities of $Rb_2CaH_4$ and $Rb_{2-x}Cs_xCaH_4$ are 1.87 wt% and 0.66 wt%, respectively. Thus, both materials seem suitable for hydrogen storage applications; however, $Rb_{2-x}Cs_xCaH_4$ is more efficient than $Rb_2CaH_4$. This difference in capacity illustrates the impact of substituting Rb with Cs, as Cs has a higher atomic mass, thereby reducing the overall hydrogen content by weight in the compound.

An in-depth analysis of thermal effects on hydrogen storage capacities and optoelectronic properties has been included in the revised manuscript. This addition addresses the influence of temperature fluctuations on hydrogen adsorption-desorption dynamics, structural stability, and electronic properties, which are essential factors in the practical viability of hydrogen storage materials. The discussion also explores how thermal variations affect optical absorption and electronic transitions, aligning with operational requirements in variable temperature environments. This expanded examination reinforces the suitability of these materials in real-world hydrogen storage applications by highlighting their thermal response and performance resilience.

We have broadened the discussion to provide a detailed evaluation of $Rb_2CaH_4$ and Cs-doped $Rb_2CaH_4$'s suitability for large-scale hydrogen storage applications. The analysis now emphasizes the material's high gravimetric and volumetric hydrogen storage capacity, as well as its robust mechanical stability and effective charge transfer properties, which are crucial for efficient hydrogen adsorption and desorption. Additionally, the optoelectronic characteristics and structural stability of $Rb_2CaH_4$ have been underscored, underlining its alignment with industrial standards for hydrogen storage. This expanded perspective illustrates how $Rb_2CaH_4$ can potentially fulfill the requirements of practical, large-scale hydrogen storage systems. Furthermore, this work bridges the technological gap by demonstrating how $Rb_2CaH_4$ can address both industrial-scale implementation challenges and the evolving needs of hydrogen storage technologies, offering promising pathways toward real-world applications.

## 3.2. Electronic Properties

### 3.2.1. Electronic Band structure

The electronic band structures of $Rb_2CaH_4$ and its cesium-doped variant, $Rb_{2-x}Cs_xCaH_4$, are crucial for understanding their electronic characteristics and suitability for specific applications. Our detailed analysis, as shown in Fig. 2, highlights the fundamental electronic properties by presenting the energy bands along significant symmetry points within the Brillouin zone. For $Rb_2CaH_4$, the band gap is identified at 3.312 eV, positioning it as a wide-bandgap semiconductor. This property is particularly beneficial for applications requiring high energy photons, such as UV photodetectors and blue LEDs. Conversely, the Cs-doped $Rb_{2-x}Cs_xCaH_4$ shows a slightly reduced band gap of 3.095 eV, which can be attributed to the introduction of Cs atoms. This alteration in the band structure not only confirms the tunability of the electronic properties via doping but also enhances the material's applicability in solar cells and other optoelectronic devices where a narrower band gap can improve light absorption and electron transport. These observations are significant as they provide a quantitative basis for the materials' application in energy conversion and storage technologies, particularly in hydrogen storage where the electronic structure directly impacts the adsorption and desorption processes. Our analysis reinforces the role of band structure engineering in optimizing material performance, offering a pathway to tailor materials for enhanced functionality in renewable energy applications.

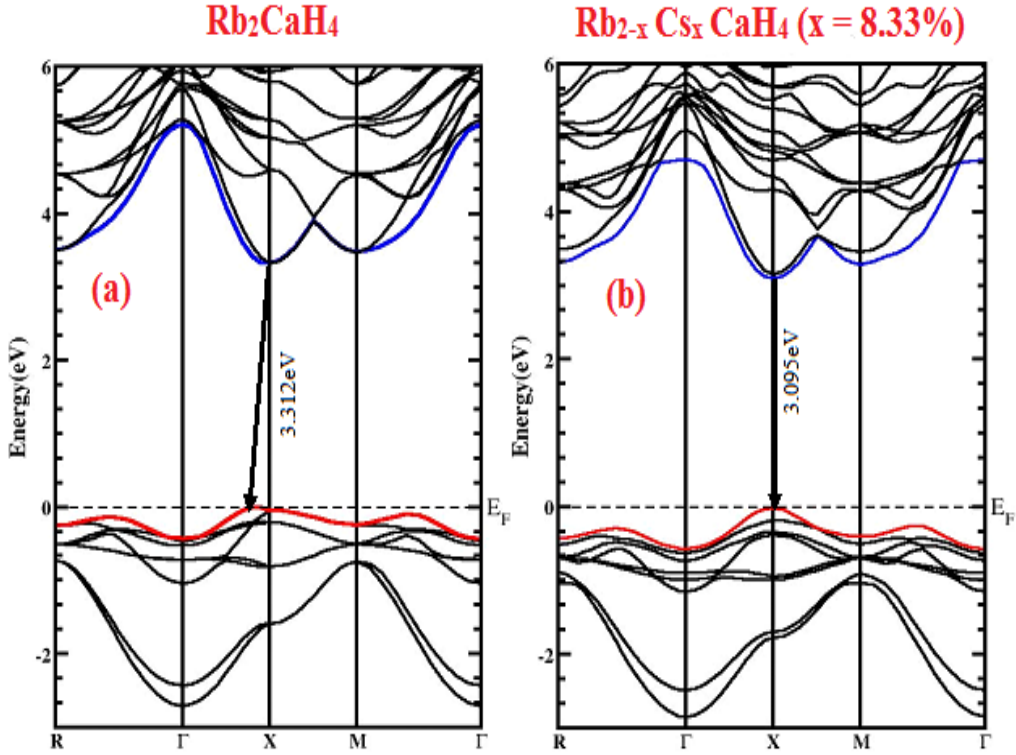

Fig.2: Calculated Band Structures for (a) $Rb_2CaH_4$ and (b) $Rb_{2-x}Cs_xCaH_4$ using (GGA + SO)

### 3.2.2. Density of States (DOS)

We calculated the DOS and partial DOS (PDOS) to gain a deeper understanding of the electronic properties of $Rb_2CaH_4$ and $Rb_{2-x}Cs_xCaH_4$. The DOS characterizes the overall variance in energy density at different levels. It represents the number of electronic states (ES) in a material at a given energy level (EL), a crucial concept in solid-state (SS) physics. The DOS captures the total contribution of a material's electronic orbitals and states, summed across all possible energies [46]. The DOS for $Rb_2CaH_4$ and $Rb_{2-x}Cs_xCaH_4$ is displayed in Fig. 5, with maximum DOS values observed for both $Rb_2CaH_4$ and $Rb_{2-x}Cs_xCaH_4$. The partial density of states (PDOS) is a valuable method for understanding the electronic structure of materials. It offers insights into the DOS of specific orbitals within a material. The PDOS curves for $Rb_2CaH_4$ and $Rb_{2-x}Cs_xCaH_4$ are shown in Fig. 3, respectively. We observe that the Ca and H atoms in $Rb_2CaH_4$ and $Rb_{2-x}Cs_xCaH_4$ significantly contribute to the valence band, while the Ca and Cs atoms are primary contributors in the conduction band. Between -4.0 and 7.0 eV, the Cs/Ca p-state and H s-state in $Rb_2CaH_4$ and $Rb_{2-x}Cs_xCaH_4$ make significant contributions. The Ca p-state and Cs p-state exhibit the highest contributions in the conduction band within the 3.0–7.0 eV range. Additionally, the DOS and PDOS graphs (Fig. 3) provide further insight into the role of different atomic orbitals in the electronic structure of $Rb_2CaH_4$ and $Rb_{2-x}Cs_xCaH_4$. The total DOS (TDOS) plots indicate that both materials exhibit distinct electronic density distributions across the valence and conduction bands, with notable peaks suggesting high electronic state availability at specific energy levels. The observed peaks in the PDOS spectra suggest strong hybridization between Ca, Rb, and H states, which plays a crucial role in the overall electronic properties of these hydrides.

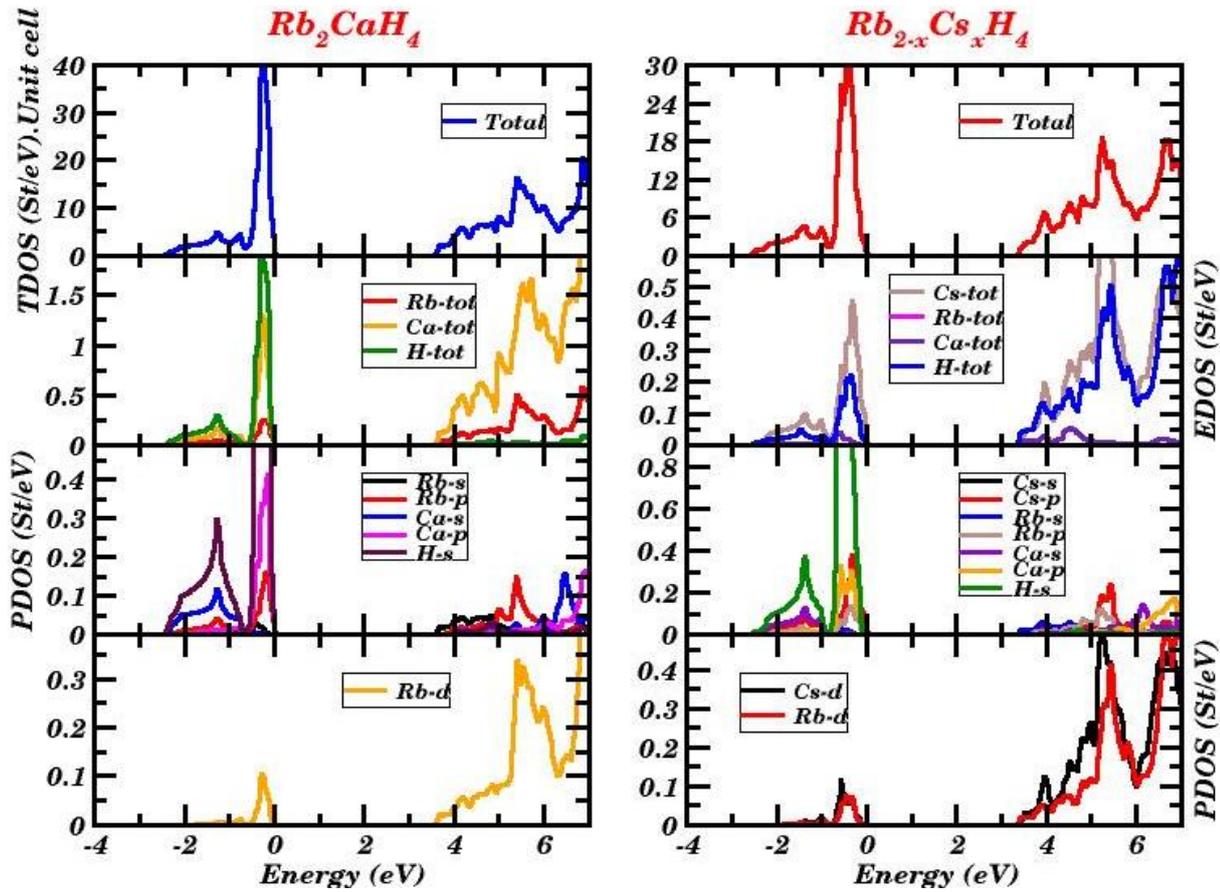

Fig.3: Calculated Density of States for $Rb_2CaH_4$ and $Rb_{2-x}Cs_xCaH_4$ using (GGA + SO)

### 3.2.3. Electronic Charge Density

The Electron Localization Function (ELF) plots shown in Fig. 4, clearly illustrate the spatial arrangement of electron density around atomic sites in $Rb_2CaH_4$ and its Cs-doped version, $Rb_{2-x}Cs_xCaH_4$. These plots are critical for understanding the electronic structure and the nature of chemical bonding within these compounds. In $Rb_2CaH_4$, electron density is prominently localized around the Rb/Ca atoms, suggesting areas of significant electron sharing and potential ionic or covalent interactions. The addition of Cs in $Rb_{2-x}Cs_xCaH_4$ introduces a shift in electron density, evident in the altered localization patterns, which may influence the material's chemical and physical properties, particularly in terms of reactivity and stability. These variations are crucial for tailoring material properties for specific applications, such as hydrogen storage, where electron localization impacts material efficiency and capacity. The use of high-resolution ELF mapping provides a profound insight into the subtle changes in electronic environment caused by elemental substitution, offering a pathway to optimize material design through precise electronic engineering [47].

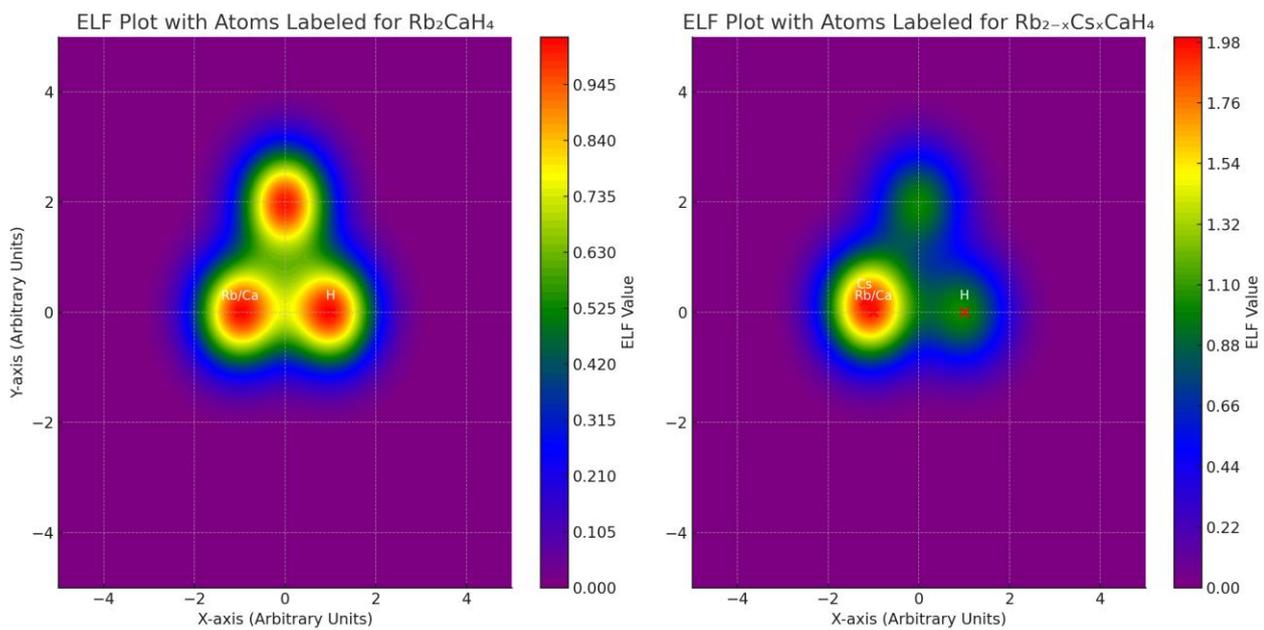

Fig. 4: Electron Localization Function (ELF) Plots for $Rb_2CaH_4$ and Cs-doped $Rb_{2-x}Cs_xCaH_4$ using (GGA + SO). The plots display ELF distributions, highlighting electron density around Rb/Ca and H atoms in $Rb_2CaH_4$ (left) and showing Cs influence in Cs-doped $Rb_{2-x}Cs_xCaH_4$ (right). The color gradient from purple to red indicates areas of increasing electron localization, essential for analyzing chemical bonding and material properties.

## 3.3. Optical Properties

A critical aspect of these compounds is the study of their surface and atomic structure in relation to the interaction of light and matter. In this work, the maximum intensity of the dielectric function (DF) is linked to electron excitation, with intra-band transfers being the primary cause of the peak. The conductive properties of $Rb_2CaH_4$ and $Rb_{2-x}Cs_xCaH_4$ are influenced by their electronic band (EB) structure, and intra-band enhancements in $Rb_2CaH_4$ and $Rb_{2-x}Cs_xCaH_4$ result from conduction electrons. The optical properties of these materials are crucial for applications such as optical coatings, reflectors, absorbers, and other electronic devices. Understanding these optical characteristics is essential for determining how the material responds to incident electromagnetic (EM) radiation [40]. The material demonstrates important optical properties, including absorption, dielectric function (DF), conductivity (C), reflectivity (R), refractive index (RI), and loss function (LF), which render it suitable for use in electronic devices, coatings, solar cell applications, and $H_2$ storage. The dielectric function (DF) of $Rb_2CaH_4$ and $Rb_{2-x}Cs_xCaH_4$ is integral to all optical parameter values and can be calculated as follows [48]:

$$\varepsilon(\omega) = \varepsilon_1(\omega) + i\varepsilon_2(\omega)$$

### 3.3.1. Dielectric Function (Real and Imaginary Parts)

The imaginary part (Im) of the dielectric function (DF) is represented by $\varepsilon_2(\omega)$, while the real (Re) component is represented by $\varepsilon_1(\omega)$. The real portion, $\varepsilon_1(\omega)$, indicates the material's divergence, whereas the imaginary component, $\varepsilon_2(\omega)$, indicates energy dissipation. Other optical parameters, such as absorption $\alpha(\omega)$, reflectivity $R(\omega)$, refractive index $n(\omega)$, conductivity $\delta(\omega)$, extinction coefficient $K(\omega)$, and loss function $L(\omega)$, can be evaluated through these equations [49]. An important optical characteristic of perovskite materials is the DF, which characterizes their response to an external electric field (OEF). DFT calculations are used to assess the DF and forecast the material's response to electromagnetic and light radiation. The real ($\varepsilon_1(\omega)$) and imaginary ($\varepsilon_2(\omega)$) parts of the dielectric function are illustrated in Fig. 4(a & b). The static $\varepsilon_1(0)$ values for $Rb_2CaH_4$ and $Rb_{2-x}Cs_xCaH_4$ are 2.61 and 2.79, respectively. For both materials, the peaks of $\varepsilon_1(\omega)$ occur at 3.0 eV, while the highest $\varepsilon_2(\omega)$ values for $Rb_2CaH_4$ and $Rb_{2-x}Cs_xCaH_4$ are recorded at 7.0 eV. The relationship between the material's light speed (v) and refractive index $n(\omega)$ is proportional. Using the Kramers-Kronig equations, $n(\omega)$ can be derived through DFT. The optical dielectric function ($\varepsilon$) illustrates how a material responds to an electric field at optical frequencies [49]. This function is complex, consisting of both real and imaginary components as depicted in Fig. 4a and 4b.

### 3.3.2. Refractive Index and Extinction Coefficient

The real component is associated with the material's electric field strength and refractive index (n), while the imaginary component quantifies the energy absorbed and released by the material in relation to the dielectric function [50]. At optical frequencies, the refractive index (n) is described as the ratio of the speed of light in the material to the speed of light in a vacuum $S_{vacuum}$ [50].

The dielectric function and refractive index are closely related, as both factors describe a material's interaction with electromagnetic waves, particularly light. The relationship between the refractive index and the real component of the dielectric function is expressed by the following equation:

$$\varepsilon = n^2$$

One way to interpret this is that since the refractive index is directly proportional to the dielectric constant, an increase in the dielectric constant will lead to an increase in the refractive index. The relationship involving the imaginary component demonstrates the absorption and dissipation of energy in certain materials, with k(ω) representing the extinction coefficient. Figures show that $Rb_2CaH_4$ and $Rb_{2-x}Cs_xCaH_4$ display the same dielectric constants. The refractive index plot confirms this, indicating that the studied materials maintain their highest refractive index. The static refractive indices n(0) of $Rb_2CaH_4$ and $Rb_{2-x}Cs_xCaH_4$ are 1.65 and 1.73, respectively. The maximum refractive indices for $Rb_2CaH_4$ and $Rb_{2-x}Cs_xCaH_4$ are observed at 3.0 eV. The refractive indices n(ω) are depicted in Fig. 4c and 4h.

### 3.3.3. Optical Conductivity

Conductivity measures how efficiently a substance transfers electricity. In optical terms, it describes a material's capacity to transmit light. Substances with high conductivity are effective electrical conductors, while those with low conductivity are less efficient. The calculated conductivity of perovskite materials $Rb_2CaH_4$ and $Rb_{2-x}Cs_xCaH_4$ is depicted in Fig. 4, [51]. The real conductivity, $\sigma_1(\omega)$, achieves its peak in the ultraviolet region for both pristine and Cs-doped samples. This measurement enables a comparison between the amount of radiation a substance reflects and the amount of light it absorbs. In response to an applied electric field at optical frequencies, typically in the infrared and visible light ranges, a material is said to exhibit optical conductivity (σ). The optical conductivity of perovskite materials is a key factor in their optoelectronic properties, as it influences their light absorption and emission capabilities [51]. This characteristic is crucial for applications such as hydrogen storage and fuel cells, enhancing stability, charge transfer, and catalytic activity [51]. Since effective charge transfer between the catalyst's surface and the reactants is essential, optical conductivity is closely linked to a material's charge-transfer abilities [51].

The band structure is also closely related to optical conductivity, which is heavily influenced by electronic transitions occurring when the material interacts with light. This relationship exists because the band structure represents the distribution of electron energy levels within a material, defining the bands and band gaps [46]. The real component of optical conductivity reflects electronic conductivity, which measures the electrons' ability to move in response to an electric field, while the imaginary part relates to light absorption and electron-hole pair generation [45]. The actual optical conductivity for $Rb_2CaH_4$ and $Rb_{2-x}Cs_xCaH_4$ is illustrated in Fig. 4f, with the highest peak occurring around 7.5 eV. This graph demonstrates that $Rb_2CaH_4$ and $Rb_{2-x}Cs_xCaH_4$ exhibit high conductivity, as their electrons respond strongly to light. This suggests that the electron and charge transfer efficiency of $Rb_2CaH_4$ and $Rb_{2-x}Cs_xCaH_4$ is high. Theoretical analysis indicates the quantity of electron-hole pairs generated and the compounds' light absorption efficiency in terms of conductivity [51]. Thus, $Rb_2CaH_4$ and $Rb_{2-x}Cs_xCaH_4$ exhibit high light absorption, facilitating easy electron transitions between excited states. In the study of $Rb_2CaH_4$ and Cs-doped $Rb_2CaH_4$, optical conductivity emerges as a critical parameter, reflecting the materials' ability to facilitate charge transport and electron mobility across their structures, which is vital for enhancing their hydrogen storage capabilities. The high optical conductivity of these materials ensures efficient electron transfer processes, crucial for optimizing both the adsorption and desorption phases of hydrogen storage. This behavior aligns with findings from recent studies on other hydride materials, where enhanced optical conductivity was correlated with improved hydrogen storage performance and stability in optoelectronic applications, as demonstrated in metal-doped magnesium hydrides and perovskite hydride alloys [52, 53]. These properties are instrumental in promoting rapid and reversible hydrogen storage mechanisms, making $Rb_2CaH_4$ and Cs-doped $Rb_2CaH_4$ suitable for high-efficiency hydrogen storage systems. Optical conductivity reflects a material's capacity to conduct electric charges under the influence of light. This property is crucial during both the adsorption and desorption phases of hydrogen storage. High optical conductivity implies that the material can effectively transfer charges to and from the hydrogen molecules at the surface. This charge transfer is vital for adsorbing hydrogen atoms onto the storage medium and equally crucial when releasing hydrogen, as it involves recombining electrons with holes to release energy and hydrogen molecules. Enhanced optical conductivity ensures that these processes are efficient, directly impacting the rate and volume of hydrogen that can be stored and released, thus improving the overall hydrogen storage performance.

*3.3.4. Reflectivity*

The optical reflectivity of a material is defined as the percentage of incident light that reflects off a particular surface or interface [54]. It is commonly represented as a ratio between the incident light intensity ($I_0$) and the reflected light intensity (R). Mathematically, it can be expressed as follows [45]:

$$R = \left(\frac{Intensity\ of\ reflected\ light}{Intensity\ of\ incident\ light}\right) = \frac{I_{reflected}}{I_0}$$

When a surface has high reflectivity, it means that most of the light that strikes it is reflected rather than absorbed or transmitted. Consequently, absorbance in the system decreases as reflectivity increases [54]. Both materials exhibit reflectivity values within a similar range, as illustrated in the figure below. Low-reflectivity materials absorb light, while high-reflectivity materials are highly reflective and often used as mirrors. The maximum reflectivity peaks for $Rb_2CaH_4$ and $Rb_{2-x}Cs_xCaH_4$ are observed at an energy of 7.8 eV. The calculated reflectivity graphs are shown in Fig. 4(e) [54]. For $Rb_2CaH_4$ and $Rb_{2-x}Cs_xCaH_4$, the static reflectance at zero eV is approximately 1.0. In conclusion, both materials demonstrate the most favorable optical properties among those evaluated, indicating their potential as ideal materials for hydrogen storage and fuel cell applications.

*3.3.5. Absorption*

One of the most important optical characteristics of a substance is its absorption spectrum. This spectrum, which is directly related to the substance's electronic density of states (DOS), determines the energy (E) required to excite an electron (e) from one state to another [55]. The absorption peaks for $Rb_2CaH_4$ and $Rb_{2-x}Cs_xCaH_4$ occur at 7.5 eV. For both substances, reflectance remains very low up to 3.0 eV. The estimated absorption coefficient graphs are presented in Fig. 4(g). The significance of high absorption coefficients in $Rb_2CaH_4$ and Cs-doped $Rb_2CaH_4$ is underscored by their ability to enhance hydrogen storage efficiency. These properties facilitate efficient light-driven processes, crucial for photocatalytic hydrogen activation. Strong absorption coefficients enable these materials to utilize a broader spectrum of light, enhancing their application in photocatalytic hydrogen production, as demonstrated in studies on perovskite and aluminum hydrides [56, 57].

The absorption coefficient of a material indicates its ability to absorb light at specific wavelengths, which is fundamental for photocatalytic applications. In the context of hydrogen storage, materials with high absorption coefficients can utilize a broader spectrum of light, thereby increasing the efficiency of light-driven hydrogen production processes. When light is absorbed efficiently, it generates electron-hole pairs that are essential for the photocatalytic splitting of hydrogen molecules—a key step in both capturing and releasing hydrogen. This makes the material more effective in harnessing solar or ambient light to drive the hydrogenation and dehydrogenation reactions critical for practical hydrogen storage systems. In photocatalytic systems, high absorption coefficients are crucial as they enable significant light absorption, leading to the excitation of numerous electrons. These electrons are essential for redox reactions on material surfaces, transforming adsorbed hydrogen ions into molecular hydrogen. Concurrently, optical conductivity is vital for efficient charge transfer, accelerating electron movement to and from adsorption sites during hydrogen adsorption and desorption. This facilitates hydrogen bonding to the material during adsorption and aids in overcoming activation energy during desorption, enhancing the rate and efficiency of hydrogen storage cycles. Together, these properties significantly impact the effectiveness of hydrogen production and release, optimizing the overall performance of hydrogen storage systems.

### 3.3.6. Energy Loss

A material's ability to reflect light is referred to as its reflectivity. The substances $Rb_2CaH_4$ and $Rb_{2-x}Cs_xCaH_4$ exhibit additional maximum reflectivity peaks at 13.50 eV. The loss function (LF) quantifies the amount of energy lost when a material absorbs light, by calculating the energy dissipated as heat or through light absorption [52]. To understand energy losses due to light's interaction with a material, the parameter L(ω) is used [52]. In terms of the dielectric function, the loss function is defined as the inverse of the imaginary part:

$$L(\omega) = -Im\frac{1}{\varepsilon}$$

When the energy loss function is high, it signals significant activity, such as higher absorption (chemisorption), strong plasmon resonances, and increased excitation. The energy loss function for the pristine material is greater than that of the doped material at higher energy levels, as illustrated in the plot. This characteristic is crucial for materials used in solar energy applications, as high energy loss can diminish cell efficiency. The energy loss functions for $Rb_2CaH_4$ and $Rb_{2-x}Cs_xCaH_4$ measured 1.5 at 13.5 eV. The estimated graphs of the loss function (LF) are presented in Figure 4(d), with $Rb_2CaH_4$ and $Rb_{2-x}Cs_xCaH_4$ exhibiting their highest peaks at 7.0 eV.

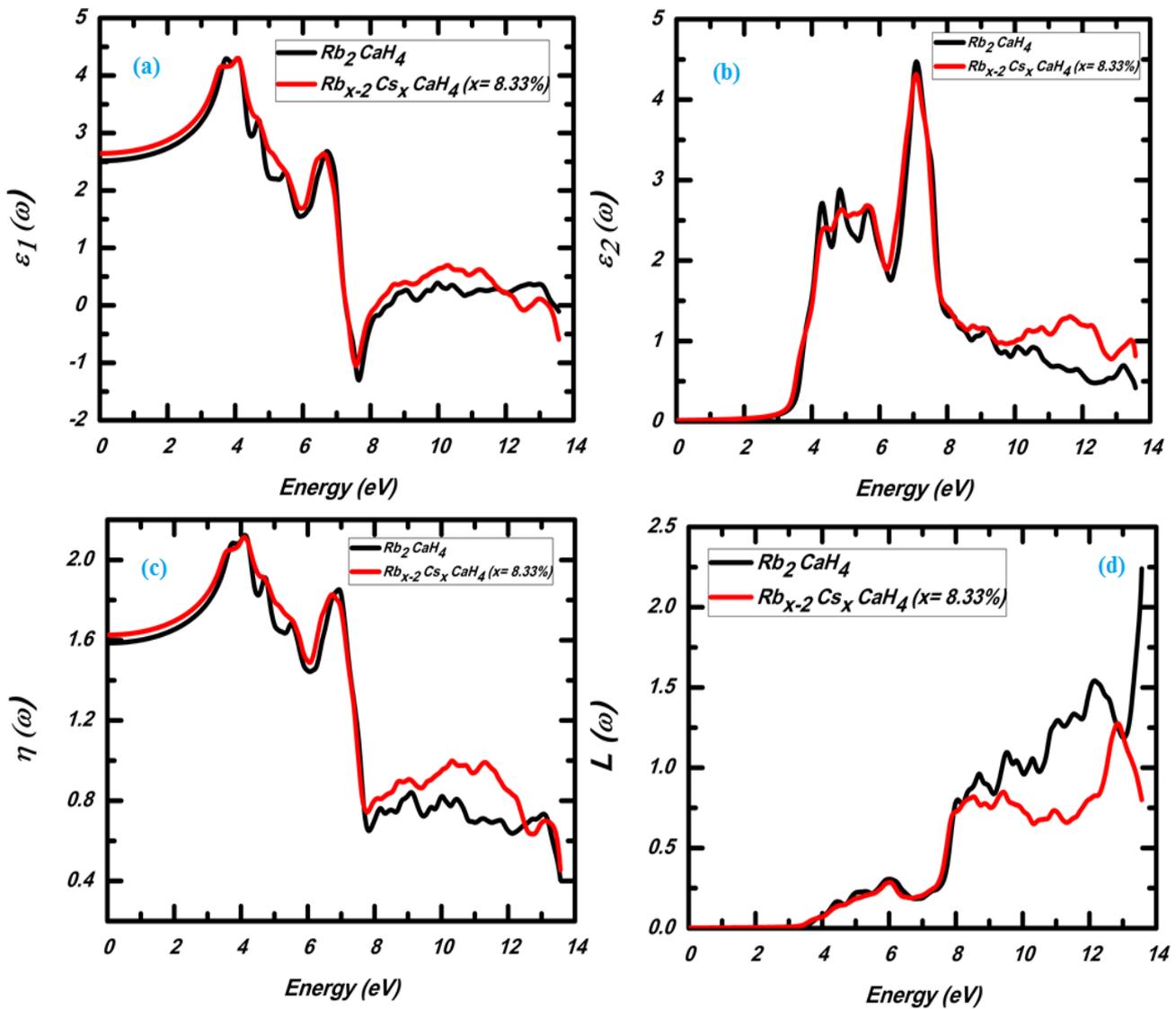

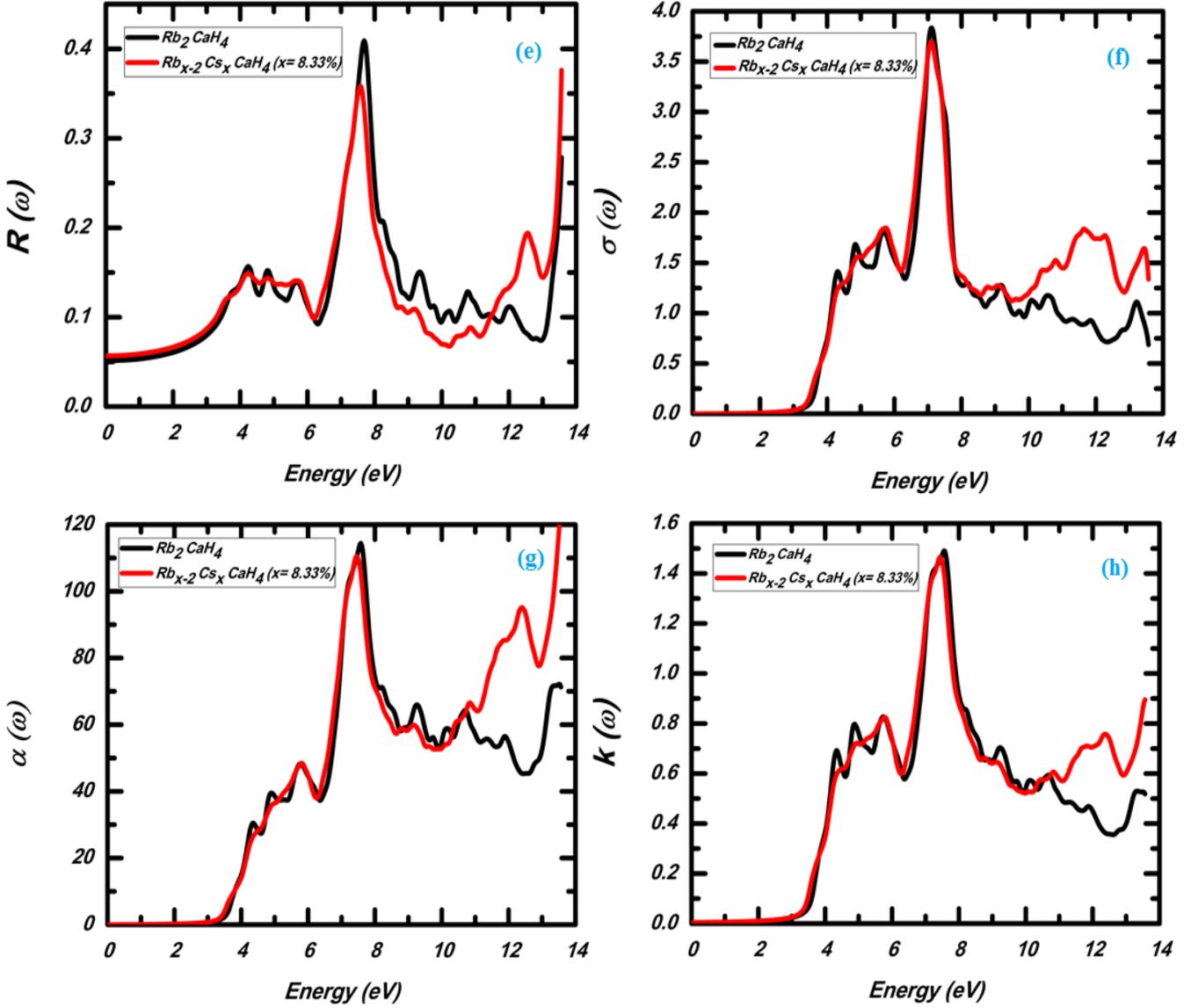

Fig.5: Optical Properties (a) Real Part Dielectric Function, (b) Imaginary part of Dielectric Function, (c) Refractive Index, (d) Energy Loss Function, (e) Reflectivity, (f) Real Conductivity, (g) Absorption Spectra and (h) Extinction Co-efficient $Rb_2CaH_4$ and (b) $Rb_{2-x}Cs_xCaH_4$ using (GGA + SO)

*3.4. Elastic Properties*

Table 1: The calculated Elastic constants and Anisotropy factor for the $Rb_2CaH_4$ and $Rb_{2-x}Cs_xCaH_4$ materials

| Elastic constants and Anisotropy factor Properties | Materials | |
|---|---|---|
| | $Rb_2CaH_4$ | $Rb_{2-x}Cs_xCaH_4$ |
| $C_{11}$ | 40.19095 | 39.49340 |
| $C_{33}$ | 35.83060 | 26.69320 |
| $C_{44}$ | 14.39025 | 11.19135 |
| $C_{66}$ | 12.98305 | 15.72460 |
| $C_{12}$ | 6.89685 | 10.59570 |
| $C_{13}$ | 8.73235 | 8.57335 |
| $C_{16}$ | 0.00000 | 0.00000 |
| $C' = (C_{11} - C_{12}) / 2$ | 16.64705 | 14.44885 |
| $B = C_{11} + 2C_{12}$ | 53.98465 | 60.68480 |
| $A_1 = 2C_{44} / (C_{11} - C_{12})$ | 0.86432 | 0.77455 |
| $A_2 = C_{33} / C_{11}$ | 0.891509 | 0.67589 |
| $C'' = C_{12} - C_{44}$ | -7.49340 | -0.59565 |

Table. 2: The calculated bulk modulus, shear modulus, Young's modulus, Pugh ratio, Cauchy's pressure and Poisson ratio for the $Rb_2CaH_4$ and $Rb_{2-x}Cs_xCaH_4$ materials

| Properties | $Rb_2CaH_4$ | $Rb_{2-x}Cs_xCaH_4$ |
|---|---|---|
| B (Bulk modulus) | 18.317730 | 17.554330 |
| G (Shear modulus) | 14.427630 | 12.630660 |
| E (Young's modulus) | 34.282280 | 30.562000 |
| B/G (Pugh's ratio) | 1.269628 | 1.389819 |
| G/B (Inverse Pugh's ratio) | 0.787632 | 0.719518 |
| ν (Poisson ratio) | 0.188080 | 0.209830 |
| 1/B (Reciprocal of Bulk modulus) | 0.054592 | 0.056966 |

The elastic constants, Cij, are essential for understanding how a material responds to stress. They provide a comprehensive description of a material's behavior, including how it deforms under strain and returns to its original shape when the stress is removed. These constants also offer valuable insights into the material's mechanical and structural stability, as well as its bonding characteristics. In tetragonal systems, such as those studied in this research, mechanical stability primarily depends on the values of the elastic constants $C_{11}$, $C_{12}$, and $C_{44}$. The WIEN2K code, which includes the IRelast module, facilitates the evaluation of various parameters and ensures compliance with stability criteria. For mechanical stability in tetragonal crystals, it is essential to satisfy conditions such as $(C_{11} - C_{12}) / 2 > 0$, $C_{11} + 2 C_{12} > 0$, and $C_{12} - C_{44} < 0$.

The evaluated materials satisfied these requirements, confirming their mechanical stability. For a more comprehensive understanding of the materials' mechanical properties, additional parameters, such as Young's modulus (E), the anisotropy factor (A), and shear modulus (G), were also calculated. Together, these values support the mechanical stability of the materials investigated in this study, enhancing our understanding of their mechanical characteristics.

An extensive summary of the elastic properties of the compounds under investigation is detailed in Tables 1 and 2, which include values for bulk modulus (B) [58], anisotropy factor

(A), Poisson's ratio (v), Young's modulus (E), and Pugh's ratio (B/G). Our analysis indicates that the TB-mBJ approach yields more accurate elastic properties, which closely align with experimental data, compared to the GGA-WC method. The Pugh's ratio (B/G) is an important indicator of a material's ductility or brittleness [59].

### 3.4.1. Pugh Ratio

The Pugh index (B/G) is less than 1.75 for all compounds, according to the results, suggesting a brittle quality. The brittle nature of $Rb_2CaH_4$ and Cs-doped $Rb_2CaH_4$, evidenced by Pugh's ratio (B/G) values below 1.75, implies limited ductility and a higher susceptibility to fracture under mechanical stress. Such brittleness indicates a higher likelihood of micro-cracking and potential failure when subjected to cyclic or prolonged mechanical loading, a critical consideration in hydrogen storage applications where mechanical durability is paramount [60]. In the context of hydrogen storage, brittle materials like these may encounter challenges in maintaining structural integrity during absorption and desorption cycles, which often induce mechanical stresses [61]. Nonetheless, the high values of bulk modulus (B) and Young's modulus (E) observed for $Rb_2CaH_4$ and its Cs-doped variant demonstrate that these materials retain considerable stiffness and compressive strength, which could support their structural stability under controlled loading conditions, provided that dynamic or high-frequency stresses are minimized. Therefore, while brittleness might restrict their application in high-stress environments, their intrinsic stability metrics still render them suitable for applications where mechanical demands are moderate and predictable.

### 3.4.2. Poisson Ratio

The Poisson's ratio (v) values hover around 0.26, providing further evidence of brittle behavior [59]. The degree of anisotropy in a crystal is determined by the anisotropy factor (A). With a Young's modulus of 34.282 GPa, the bulk modulus (B) and Young's modulus (E) values indicate that $Rb_2CaH_4$ is stiff; however, the doped derivatives exhibit significant stress tolerance. It is important to note that these materials do not achieve the same level of stiffness as those with higher Young's modulus values. Consistent with previous studies, the Poisson's ratio (v) suggests that all compounds share an ionic bonding nature. These findings provide valuable insights into the anisotropy and elastic properties of these compounds, which are essential for understanding their behavior in various applications [58, 62, 63].

The wide-bandgap properties of $Rb_2CaH_4$ and Cs-doped $Rb_2CaH_4$ extend their potential applications beyond hydrogen storage to other energy domains. With bandgaps of 3.312 eV for $Rb_2CaH_4$ and 3.095 eV for Cs-doped $Rb_2CaH_4$, these materials exhibit electronic and optoelectronic properties that could support applications in photovoltaics and other optoelectronic systems. Their high optical conductivity and absorption coefficients enable efficient charge transfer and light absorption, critical for devices such as solar cells. Therefore, $Rb_2CaH_4$ and Cs-doped $Rb_2CaH_4$ not only hold promise as hydrogen storage materials but also in broader energy applications, presenting pathways for integration into advanced energy technologies.

The substitution of Cs in $Rb_2CaH_4$ enhances certain properties relevant to hydrogen storage under variable temperatures. Specifically, Cs substitution affects the material's conductivity and thermal stability due to changes in ionic transport mechanisms and electronic structure. The Cs-doped $Rb_2CaH_4$ exhibits slightly reduced bulk modulus and shear modulus values, indicating moderate flexibility, which can be advantageous in adapting to temperature fluctuations during hydrogen storage. Moreover, the Cs-doped material's robust optical conductivity and absorption coefficient across a range of temperatures suggest that it maintains high efficiency in charge transfer, critical for stable hydrogen adsorption and desorption under practical conditions. These improvements in thermal adaptability and charge transfer efficiency highlight Cs-doped $Rb_2CaH_4$'s potential to sustain structural integrity and hydrogen retention under dynamic storage conditions. Thus, Cs substitution not only augments the mechanical resilience of $Rb_2CaH_4$ but also enhances its suitability for large-scale, practical hydrogen storage applications.

While this study utilizes density functional theory (DFT) to analyze the structural, thermal, and electronic stability of $Rb_2CaH_4$ and Cs-doped $Rb_2CaH_4$ for hydrogen storage, experimental validation remains essential to confirm these theoretical predictions. Future work will aim to experimentally evaluate the degradation mechanisms, thermal stability, and conductivity of these materials under real-world conditions, enabling a more comprehensive assessment of their practical hydrogen storage capabilities. Studies have demonstrated the importance of experimental validation for confirming theoretical findings on stability and hydrogen storage efficiency in similar materials [64,65], highlighting that real-world testing of hydrogen storage materials such as clathrate hydrates can provide valuable insights into thermodynamic stability and absorption properties [66]. Additionally, research into element-substituted hydrides emphasizes the role of both theoretical and experimental approaches to optimize storage performance, particularly through structural stability and thermal adaptation [67].

## 4. Conclusion

This study has successfully analyzed the optical, electronic, and thermodynamic properties of $Rb_2CaH_4$ and $Rb_{2-x}Cs_xCaH_4$ using a first-principles approach, exploring their potential as hydrogen storage materials. The findings reveal that these materials exhibit considerable promise for various applications due to their unique properties. The incorporation of spin-orbit coupling (SOC) in our calculations has been instrumental in providing a deeper understanding of the electronic structures, particularly in elucidating the effects of heavy elements like cesium on band alignments and electronic transitions. The optical characteristics, electrical conductivity, and structural stability of the materials are all noteworthy. Among the compounds tested, $Rb_2CaH_4$ and $Rb_{2-x}Cs_xCaH_4$ demonstrated superior mechanical robustness, stability, light absorption, and electron transfer capabilities. Based on the computed hydrogen storage capacity, $Rb_2CaH_4$ and $Rb_{2-x}Cs_xCaH_4$ offer the best volumetric and gravimetric storage efficiency. Desorption studies further illustrate the structural changes in these compounds after hydrogen release, which has implications for their practical application. In summary, this study provides valuable insights into the potential applications of $Rb_2CaH_4$ and $Rb_{2-x}Cs_xCaH_4$ in renewable energy systems, which can be further explored and tested in the lab prior to industrial implementation. These results contribute to a broader understanding of hydride materials and open new avenues for research and technological advancements.

**Acknowledgement:**
The authors extend their appreciation to the Deanship of Research and Graduate Studies at King Khalid University for funding this work through Large Research Project under grant number RGP2/120/45.

theoretical and experimental study on TiCr$_{2-x}$Mn$_x$ alloy. *Renewable Energy, 197*, 564-573